\documentclass[a4,11pt, fleqn]{article}
\pdfoutput=1
\usepackage[T1]{fontenc}
\usepackage{setspace}
\usepackage{tabulary}
\usepackage{caption}
\usepackage{subcaption}
\usepackage{multirow}
\usepackage{jheppub}

\title{Using MiniBooNE neutral current elastic cross section results to constrain 3+1 sterile neutrino models}
\author[1]{Callum Wilkinson,}
\note{Corresponding author.}
\author{Susan Cartwright}
\author{and Lee Thompson}
\affiliation{Department of Physics and Astronomy, University of Sheffield,\\Hicks Building, Hounsfield Road, Sheffield, S3 7RH, United Kingdom}

\emailAdd{callum.wilkinson@sheffield.ac.uk}

\abstract{\noindent The MiniBooNE Neutral Current Elastic (NCEL) cross section results are used to extract limits in the $\Delta m^{2}-\sin^{2}\vartheta_{\mu s}$ plane for a 3+1 sterile neutrino model with a mass splitting $0.1 \leq \Delta m^{2} \leq 10.0$ eV$^{2}$. GENIE is used with a cross section model close to the one employed by MiniBooNE to make event rate predictions using simulations on the MiniBooNE target material CH$_{2}$. The axial mass is a free parameter in all fits. Sterile modifications to the flux and changes to the cross section in the simulation relate the two and allow limits to be set on sterile neutrino mixing using cross section results. The large axial mass problem makes it necessary for experiments to perform their own axial mass fits, but a prior fit to the same dataset could mask a sterile oscillation signal if the sterile and cross section model parameters are not independent. We find that for the NCEL dataset there are significant correlations between the sterile and cross 
section model parameters, making a fit to both models simultaneously necessary to get robust results. Failure to do this results in stronger than warranted limits on the sterile parameters. The general problems that the current uncertainty on charged-current quasi-elastic (CCQE) and NCEL cross sections at MiniBooNE energies pose for sterile neutrino measurements are discussed.}

\notoc
\begin{document}
\maketitle
\flushbottom

\section{Introduction}
\label{sec:intro}
There are three neutrino flavours present in the Standard Model, and further neutrinos which interact weakly are ruled out by the measurement of the Z$^{0}$ invisible decay width from a combination of LEP experiments and groups~\cite{lep06}. However, since the results of the LSND short-baseline neutrino oscillation experiment~\cite{lsnd01} there has been a great deal of theoretical interest in additional, sterile (non-weakly interacting), neutrinos with masses on the eV scale which participate in neutrino mixing. Apart from cosmological bounds~\cite{wmap11, wmap13, planck13} which have some dependence on the assumed cosmological model and the physics of neutrino production in the early universe, the experimental signature of sterile neutrinos is anomalous effects over shorter baselines than can be explained by standard three neutrino mixing, and a number of more recent experimental hints have fuelled interest in this area (for a complete review see~\cite{abazajian12}; for recent global fits to sterile 
neutrino models see~\cite{kopp13GF, conrad12GF}). 

The primary aim of this analysis is to use the published MiniBooNE Neutral Current Elastic (NCEL) cross section results~\cite{mbNCEL} to produce limits on muon to sterile neutrino mixing (limits in the $\Delta m^{2}$ - sin$^2 2\vartheta_{\mu s}$ plane) for a sterile neutrino model with a single additional mass splitting $0.1 \leq \Delta m^{2} \leq 10.0$ eV$^{2}$ (a 3+1 model). Although using cross section measurements in the context of sterile neutrinos is unusual, limits have previously been produced from $\nu_{e}-$carbon cross section measurements~\cite{conradXSec} by comparing the published results with theoretical cross section predictions. This analysis uses the GENIE interation generator~\cite{genieMC} and the relativistic Fermi gas (RFG) model of Llewellyn-Smith~\cite{smith72} to make event rate predictions with simple Monte Carlo simulations on the MiniBooNE detector medium, CH$_{2}$. The cross section model employed here closely follows the MiniBooNE cross section model~\cite{KatoriThesis, 
PerevalovThesis} and is described in detail in Section~\ref{sec:xsecModel}. One cross section parameter, the effective axial mass $M_{A}^{eff}$, is used in the fit.

The RFG model cannot adequately describe the global dataset, recent measurements of the axial mass by MiniBooNE~\cite{mbCCQE, mbNCEL, mbAntiCCQE, mbAntiNCEL}, K2K~\cite{k2kCCQE} and MINOS~\cite{minosCCQE} are incompatible with historical measurements~\cite{zeller12, ankowski12, sobczyk10}. Many papers have been written advancing various theoretical models which try to explain these differences~\cite{martini09, martini10, nieves11, amaro10, bodekTEM, butkevichCCQE, leitner08} (for a good review~\cite{zeller12}). The axial mass values measured by recent experiments must be treated as effective axial mass values, where the model still fits the data reasonably well in isolation, but it is understood that $M_{A}$ has been inflated to include various other contributions, arising from the size of the nuclear target. This effective axial mass is here denoted $M_{A}^{eff}$ to highlight this throughout. The MiniBooNE NCEL dataset has been fit to various other models which try to account for these effects in a more 
rigorous way~\cite{butkevichNCEL, sobczykNCEL, ankowski12}. In general it is not possible to take $M_{A}^{eff}$ measured by an experiment and apply it to another experiment, as the additional contributions depends on the type of target, the type of detector, and the energy distribution of the neutrino beam. For this reason, $M_{A}^{eff}$ measurements from other experiments cannot be included to constrain $M_{A}^{eff}$ in the fit. It is also not possible to include other samples to constrain the sterile mixing parameters, as this would rely on an inconsistent cross section model. Although it has been argued that large $M_{A}^{eff}$ values~\cite{mbAntiNCEL} still fit the data reasonably well, there is no current consensus on how to correctly model the cross section enhancement of a multi-nucleon target. Until this is resolved, any attempt to fit a sterile neutrino model to datasets from multiple experiments will be extremely difficult. Despite its shortcomings, the RFG model is the appropriate choice of cross 
section in this analysis because it is still the underlying model in the simulations used by the current 
generation of neutrino experiments including MiniBooNE, and therefore is commonly used to produce sterile neutrino limits.

Because each experiment must rely on its own $M_{A}^{eff}$ measurement, experiments that produce sterile neutrino limits run the risk of fitting to the same dataset twice if the cross section parameters are not varied in the fit. Current sterile limits have been produced which rely on a pre-measured value of $M_{A}^{eff}$~\cite{mbSciNu2012, mbSciaNu2012}, which is only valid if all of the fitted cross section parameters are independent of all of the sterile neutrino parameters. MiniBooNE state that their cross section and sterile parameters are uncorrelated for the $\nu_{\mu}$- disappearance measurement using their CCQE dataset~\cite{mbDisApp09}, however this may not be the case for other datasets. In this analysis, we mimic this `sequential' fitting, as well as fitting to all parameters concurrently in the `simultaneous' fit. We find that for the NCEL dataset, the sequential and simultaneous fits tend to very different best fit values, and produce very different limits, so it is important to stress in the 
introduction that only the simultaneous fit is statistically justified in the MiniBooNE NCEL case. As sequential fits have been used in the past, we include the comparison to advise caution for other sterile fits. If the fitted cross section and sterile neutrino parameters are correlated, then a sequential type fit risks masking, or partially masking, a sterile neutrino signal, or any statistical fluctuations that mimic a signal, resulting in stronger than justified limits on sterile mixing parameters. It has been pointed out in~\cite{ankowski12} that underestimated cross sections might lead to false oscillation signals - overestimating the cross section could hide an oscillation signal.

Section~\ref{sec:sterileModel} gives a brief introduction to the sterile neutrino model. Section~\ref{sec:method} gives details on how predicted distributions were produced for this analysis, including details of the cross section model and relevant details of the MiniBooNE experiment. The fit information and results are presented in Section~\ref{sec:ncelFit}, and are discussed in Section~\ref{sec:conclusion} which also contains concluding remarks.

\section{3+1 neutrino mixing}
\label{sec:sterileModel}
Sterile neutrino models that include a single additional, predominantly sterile, neutrino mass state $\nu_{4}$, which is heavier than the other, predominantly active mass states are generically referred to as 3+1 models. In this analysis, only 3+1 models were considered. It is common in the literature to refer to ``short baseline oscillations'' in the context of sterile neutrino searches~\cite{giuntiSBL11}: more precisely, this refers to oscillations where the L/E is such that standard three flavour mixing can be neglected, so any oscillations are driven by the additional, predominantly sterile, mass state. In this approximation, $\Delta m^{2}_{21} = \Delta m^{2}_{32} = \Delta m^{2}_{31} = 0$, which leaves a single mass splitting (for a 3+1 model), here denoted as $\Delta m^{2}_{42}$.

In the short baseline approximation, the appearance and disappearance probabilities are given by Equation~\ref{eq:3+1App} and Equation~\ref{eq:3+1Dis} respectively~\cite{giunti11}:
\begin{align}
\indent
P_{ \substack{(-) \\ \nu_{\alpha} }  \rightarrow \substack{(-) \\ \nu_{\beta} }}=\sin^{2} 2\vartheta_{\alpha\beta}  \sin^{2} \left(\frac{1.265\Delta m^{2}_{42} [\mathrm{eV}^{2}] L [\mathrm{km}]}{\
E [\mathrm{GeV}]} \right ) , && \left(\alpha\neq\beta\right)
\label{eq:3+1App}
\end{align}
\begin{align}
\indent
P_{ \substack{(-) \\ \nu_{\alpha} } \rightarrow \substack{(-) \\ \nu_{\alpha} }}=1 - \sin^{2} 2\vartheta_{\alpha\alpha}  \sin^{2} \left(\frac{1.265\Delta m^{2}_{42}L}{E} \right ) ,
\label{eq:3+1Dis}
\end{align}
\noindent for $\alpha , \beta = e, \mu , \tau , s ,$ with:
\begin{equation}
\indent
\sin^{2} 2\vartheta_{\alpha\beta}  = 4 |U_{\alpha 4}|^{2} |U_{\beta 4} |^{2} ,
\label{eq:3+1TransSin}
\end{equation}
\begin{equation}
\indent
\sin^{2} 2\vartheta_{\alpha\alpha}  = 4 |U_{\alpha 4}|^{2} \left(1 -|U_{\alpha 4} |^{2} \right).
\label{eq:3+1AppSin}
\end{equation}

Neutral Current (NC) disappearance in a purely $\nu_{\mu}$ beam can be expressed as in Equation~\ref{eq:3+1NCdis}, where the unitarity constraint $1 = |U_{e4}|^{2} + |U_{\mu 4}|^{2} + |U_{\tau 4}|^{2} + |U_{s4}|^{2}$~\cite{giunti11} has been used. As the NC signal is sensitive to $|U_{e4}|^{2} + |U_{\mu 4}|^{2} + |U_{\tau 4}|^{2}$, NC disappearance experiments can place limits in the $\Delta m^{2}-\sin^{2}2\vartheta_{\mu s}$ plane. Similiar limits would require an ensemble of charged current measurements.
\begin{align}
\indent
\notag P^{NC} &= 1 - \sin^{2} 2\vartheta_{\mu s}  \sin^{2} \left(\frac{1.265\Delta m^{2}_{42}L}{E} \right ) \\
&= 1 - 4|U_{\mu4}|^{2} \left(1 - |U_{e4}|^{2} - |U_{\mu 4}|^{2} - |U_{\tau 4}|^{2} \right) \sin^{2}\left(\frac{1.265\Delta m^{2}_{42} L}{E}\right) 
\label{eq:3+1NCdis}
\end{align}
It is known from other experimental results~\cite{giunti11}, that $U_{s4} >> U_{e4}, U_{\mu 4}, U_{\tau 4}$, otherwise all conventional neutrino experiments would have seen significant anomalies in their results. However, an NC only search may not reflect this: a large value for $U_{\mu 4}$ can be compensated by a large value of $U_{\tau 4}$ or $U_{e4}$ as the signal does not distinguish between the three active neutrino species.

Note that in this analysis, the signal comes from $\nu_{\mu}$ and $\nu_{e}$ (from contamination in the beam). For this reason, it is not sufficient to fit only to $U_{\mu 4}$ and $U_{s4}$, as $\nu_{e}$ oscillation must also be considered.

\section{Analysis method}
\label{sec:method}
\subsection{Cross section model}
\label{sec:xsecModel}
The MiniBooNE NCEL cross section results are given in terms of reconstructed kinematic variables so that theorists can use them to test different cross section models. To test oscillation hypotheses using these results, a cross section model is required to relate the energy of the incoming neutrinos to the measured kinematic variables: oscillation results are dependent on the choice of cross section model. This analysis uses a cross section model based on the RFG nuclear model from Bodek and Ritchie~\cite{bodek81} to simulate events on CH$_{2}$ (the MiniBooNE target material) using the GENIE interaction generator~\cite{genieMC}. Although MiniBooNE and this analysis use different interaction generators with different cross section parameters, the aim of this analysis is to reproduce the MiniBooNE model as closely as possible, so cross section parameters were chosen to minimise the effect of using different generators. RFG models are widely used by the current generation of neutrino experiments, and in the 
calculation of sterile neutrino limits~\cite{mbSciNu2012, mbSciaNu2012, mbDisApp09}, which makes the model an appropriate choice for this analysis.

MiniBooNE use Nuance v3~\cite{nuanceMC} as their interaction generator, which uses the Llewellyn-Smith model~\cite{smith72} to describe NCEL scattering off free protons and the Smith-Moniz model~\cite{smith-moniz72} to describe NCEL scattering off bound nucleons. In this analysis, GENIE 2.6.2~\cite{genieMC} was used, which models NCEL scattering with the formalism described by Ahrens {\it et al.}~\cite{e734NCEL}. Although BBA-03~\cite{bba03} form factors could have been used in this analysis, BBBA-05~\cite{bbba05} are the default in GENIE, reflecting a wider usage of the newer form factors, so were retained for this analysis. The MiniBooNE cross section model had a value of the strange quark contribution to the nucleon spin, $\Delta s = 0.0$, they make a measurement of this parameter in~\cite{mbNCEL} of $\Delta s = 0.08 \pm 0.26$ which they point out is in agreement with the value measured by the BNL E734 experiment~\cite{e734NCEL}, which is the GENIE default value, again used for this analysis. A summary of 
the cross section models used by MiniBooNE and in this analysis is given in Table~\ref{tab:xsec_summary}; further details for the MiniBooNE model can be found in~\cite{KatoriThesis, PerevalovThesis} from which the summary here has been drawn. $M_{A}^{eff}$ has not been included in Table~\ref{tab:xsec_summary} because the value is obtained in the fit.
\begin{table}[h]
  \centering
    \begin{tabulary}{0.8\textwidth}{CCC}
	  \hline\hline
	  & NUANCE & GENIE \\ \hline
	  Binding Energy for Carbon & 34.0 GeV & 34.0 GeV \\
	  Fermi Momentum in Carbon & 220.0 MeV & 221.0 MeV \\
	  Vector Mass, M$_{V}$ & 850 MeV & 850 MeV \\
	  Pseudoscalar Form Factors & BBA-03~\cite{bba03} & BBBA-05~\cite{bbba05} \\
	  sin$^2 \theta_{W}$ & 0.2315 & 0.2315 \\
	  Pauli Blocking, $\kappa$ & 1.0220 & 1.0 \\
	  $\Delta$s & 0.0 & -0.15 \\
	  \hline\hline	
      \end{tabulary}
  \caption{Summary of cross section parameters used in the MiniBooNE analysis (Nuance) and this analysis (GENIE).}	
  \label{tab:xsec_summary}
\end{table}
While most axial mass measurements use shape only fits~\cite{zeller12}, including the normalisation uncertainty is important for the sterile neutrino fits in this analysis, so it would have been inconsistent to omit the normalisation uncertainty from the cross section fit. As such, care should be taken when making comparisons between these results and others published elsewhere.

\subsection{Experimental details}
The signal definition and experimental details relevant for this analysis are given in Table~\ref{tab:sampleSummary}, and along with the flux prediction~\cite{mbFlux}, are all of the details required to predict the true event rate in MiniBooNE for any given sterile hypothesis.
\begin{table}[h]
  \centering
    \begin{tabulary}{1.0\textwidth}{CC}
	  \hline\hline
	  {\bf Property} & {\bf MiniBooNE NCEL} \\\hline\hline
	  Baseline L (m) & 541 \\
	  Average Neutrino Energy (GeV) & 0.788\\
	  Energy Range for Measurement (GeV) & $0 \leq E_{\nu} \leq 10$ \\
	  Signal Events & $\nu_{\mu, e} + n,p \rightarrow \nu_{\mu, e} + n,p$\\
	  POT & $6.46 \times 10^{20}$ \\
	  Integrated Flux $\Phi_{\nu}$ ($\nu$ cm$^{-2}$ POT$^{-1}$) & $5.22227 \times 10^{-10}$\\
	  Target Material & CH$_{2}$ \\
	  \hline\hline	
      \end{tabulary}
  \caption{Summary of the important experimental details for the two samples used in this analysis. Further details describing the MiniBooNE NCEL sample can be found in~\cite{mbNCEL, PerevalovThesis}.}	
  \label{tab:sampleSummary}
\end{table}

The MiniBooNE NCEL results are given as event rates in bins of T$_{reco}$, the sum of the kinetic energy of final state nucleons, for which the full covariance matrix has been provided~\cite{mbNCEL}. The MiniBooNE estimation of the beam related, and beam unrelated backgrounds are available with the T$_{reco}$ results, and were used in this analysis (details are in Section~\ref{sec:samples}). It is important to note that the effect of sterile neutrinos on the beam related backgrounds were not taken into account as there were insufficient available details to do so.

\subsection{Generating samples}
\label{sec:samples}
To perform this analysis, it was necessary to vary the cross section and sterile model parameters simultaneously in a fit, in a computationally feasible way.

GENIE provides tools for reweighting cross section parameters in a simulated sample, allowing a range of $M_{A}^{eff}$ values to be investigated using a single sample at fixed $M_{A}^{eff}$. By binning the weighted events into the desired kinematic variables, a plot of expected event rate per bin against the cross section parameter can be produced, which can then be interpolated to give a predicted event rate in each bin for any value of the cross section parameter in the range specified. For further details on event reweighting, refer to the GENIE documentation~\cite{genieMC} (also the information on the GENIE webpages).

MiniBooNE provide detailed flux information~\cite{mbFlux}, so it is trivial to produce the expected MiniBooNE flux under any sterile hypothesis by applying the equations in Section~\ref{sec:sterileModel}. Producing a predicted event rate in terms of kinematic variables from a predicted flux requires a migration matrix, where events are split into $(E_{\nu}, T_{true})$ bins. By producing a sample with a flat flux distribution, it is possible to produce an expected event rate for any sterile hypothesis using the following method, where $i$ denotes $E_{\nu}$ bins, and $j$ denotes $T_{true}$ bins:

\begin{enumerate}
 \item A two-dimensional histogram of signal events, S, with $(E_{\nu}, T_{true})$ bins was produced,
 \item A one-dimensional histogram of all simulated events, R, with $E_{\nu}$ bins was produced,
 \item A plot of the total cross section on the target molecule (CH$_{2}$) in $E_{\nu}$ bins was produced, giving $\sigma^{total}_{i}$ for all $i$,
 \item A modified flux histogram for the sterile hypothesis, $\Phi$, was produced,
 \item A scaling factor, $\epsilon_{i}$ was found for each energy bin $i$ such that $R_{i} \times \epsilon_{i} = \sigma^{total}_{i}$,
 \item The scaling factor $\epsilon_{i}$ was applied to $S_{i}$ for all $j$,
 \item $\Phi_{i}$ was multiplied by $S_{i}$ for all $j$,
 \item $S$ was projected onto the axis $j$, giving the expected event rate in terms of $T_{true}$.
\end{enumerate}

\begin{figure}[ht]
 \centering
  \includegraphics[width=0.75\textwidth]{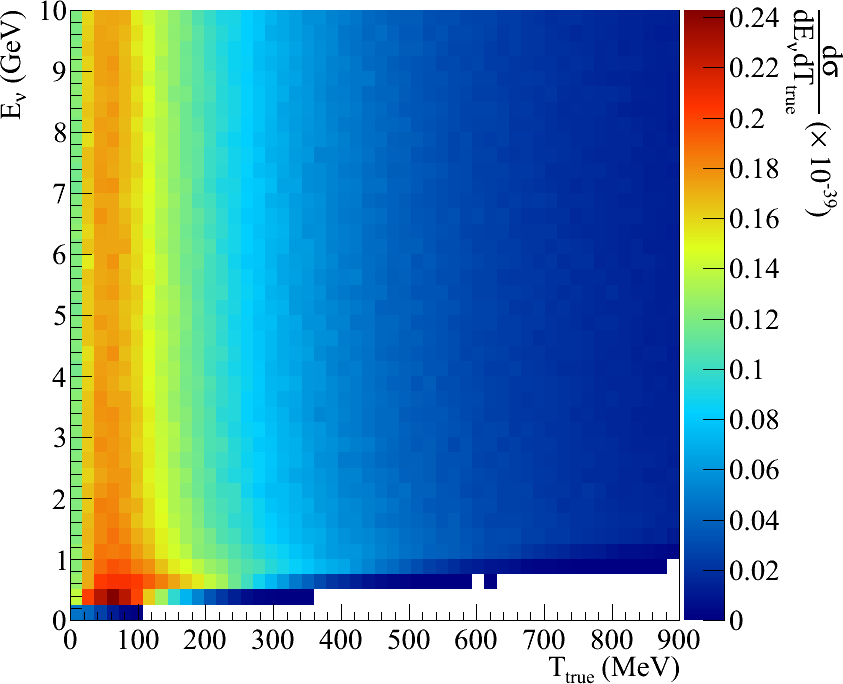}
  \caption{An example migration matrix of cross section values for the NCEL prediction in $(E_{\nu}, T_{true})$ bins for $M_{A}=1.24$ GeV.}
  \label{fig:migMatrix}	
\end{figure}

Steps 1-6 are calculated before fitting, which leaves $S$ as a matrix of cross section values for each $(E_{\nu}, T_{true})$ bin. Figure~\ref{fig:migMatrix} shows an example matrix, showing the cross section values for $M_{A} = 1.24$ GeV. Steps 7 and 8 are performed for each iteration of the fit, thus producing a predicted event rate in terms of the true values of the kinematic variables for each sterile hypothesis without having to produce a new sample at each iteration. Very large samples of $1\times 10^{8}$ neutrino interactions were produced for each neutrino flavour with a flat flux with $0 \leq E_{\nu} \leq 10$ GeV. Although computationally expensive, this was necessary to render the statistical error in the simulated samples negligible.

\begin{figure}[ht]
 \centering
  \includegraphics[width=0.9\textwidth]{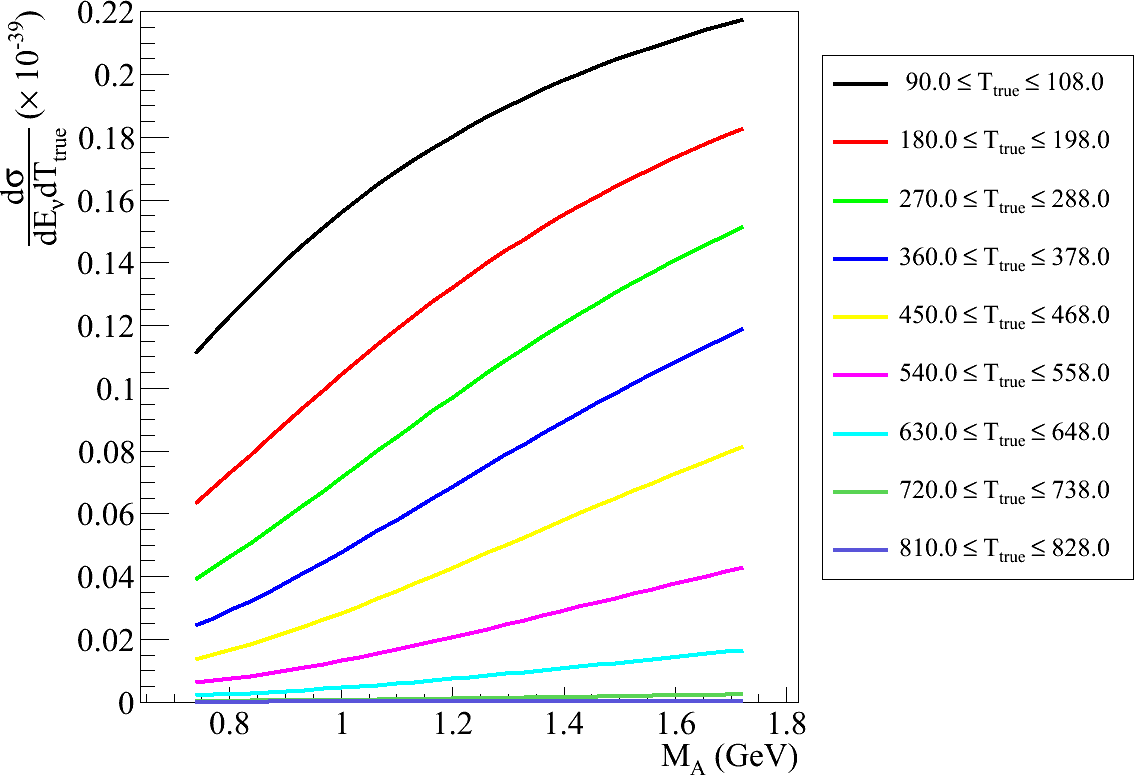}
  \caption{An example migration matrix of cross section values for the NCEL prediction in $(E_{\nu}, T_{true})$ bins for $M_{A}=1.24$ GeV.}
  \label{fig:maSplines}	
\end{figure}

It is trivial to combine the two methods described above to allow both the cross section and sterile model parameters to be varied in the fit: $S$ becomes a matrix of cross section splines rather than a matrix of cross section values. At each iteration of the fit, a matrix of values is produced by interpolating the cross section splines in each bin of $S$ to give a matrix of cross section values which can then be dealt with as described in steps 7 and 8 above. Figure~\ref{fig:maSplines} shows example cross section splines from the NCEL matrix.

The method so far gives an event rate in terms of the \emph{true} nucleon kinetic energy, T$_{true}$, whereas the NCEL results are given in terms of the \emph{reconstructed} nucleon kinetic energy, T$_{reco}$, without removing the energy smearing and detector inefficiencies. To produce an expected event rate in T$_{reco}$, it is necessary to transform the T$_{true}$ prediction using a response matrix which simulates the detector inefficiencies and energy smearing. Appendix B of~\cite{PerevalovThesis} gives all of the necessary details to use the information released with~\cite{mbNCEL} to produce a response matrix.

The GENIE simulation used in this analysis simulates all potential signal events identified in~\cite{PerevalovThesis} apart from the irreducible backgrounds. A combined response matrix for the simulated signal events is calculated as described, and used to transform the predicted T$_{true}$ event rate into T$_{reco}$ at each iteration of the fit. The T$_{reco}$ event rate distribution from the irreducible backgrounds and the beam unrelated backgrounds is added to produce a final T$_{reco}$ distribution which can be predicted with the published MiniBooNE results. It should be stressed that the beam unrelated, and more importantly, the irreducible beam related background event rates are both MiniBooNE calculations which use the MiniBooNE cross section model, not the GENIE model used for the signal events in this analysis. 

\subsection{Example plots}
\begin{figure}
 \centering
 \begin{subfigure}[h]{0.49\textwidth}
    \includegraphics[width=\textwidth]{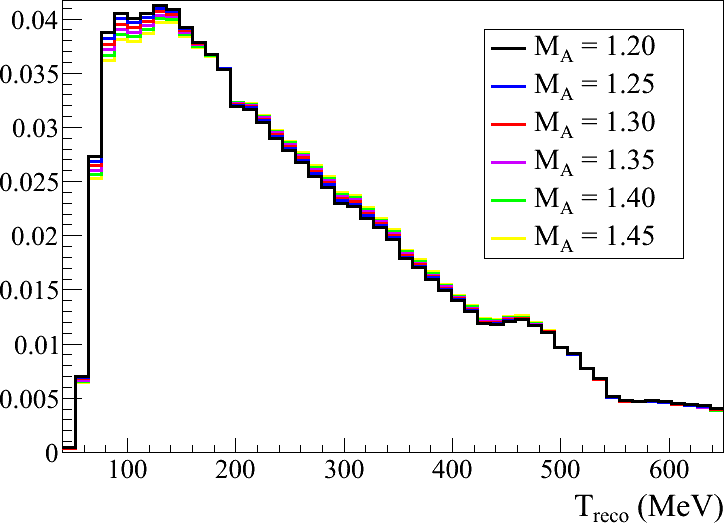}
    \caption{$M_{A}^{eff}$ shape only}
    \label{subfig:NCEL_ma_shape}
 \end{subfigure}
 \begin{subfigure}[htb]{0.49\textwidth}
    \includegraphics[width=\textwidth]{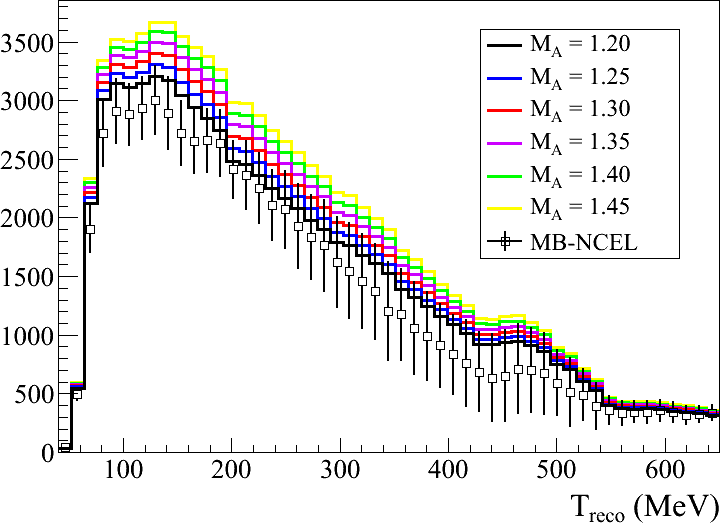}
    \caption{$M_{A}^{eff}$ with normalisation}
    \label{subfig:NCEL_ma_norm}
\end{subfigure}
 \begin{subfigure}[h]{0.49\textwidth}
    \includegraphics[width=\textwidth]{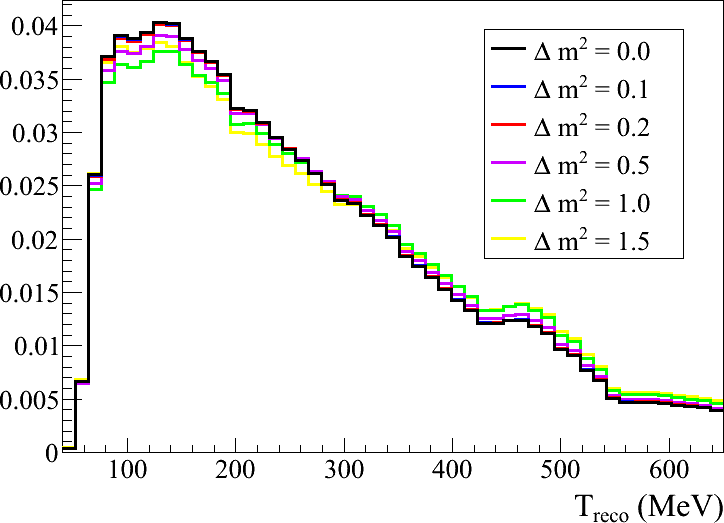}
    \caption{$\Delta m^{2}$ shape only}
    \label{subfig:NCEL_sterile_shape}	
 \end{subfigure}
 \begin{subfigure}[htb]{0.49\textwidth}
    \includegraphics[width=\textwidth]{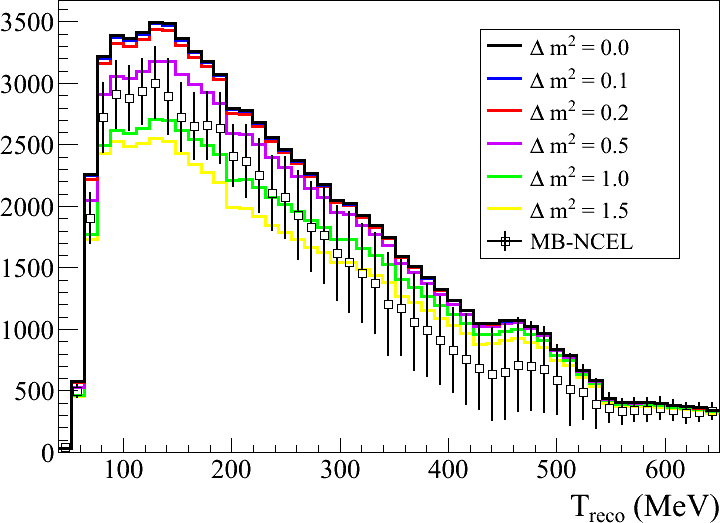}
    \caption{$\Delta m^{2}$ with normalisation}
    \label{subfig:NCEL_sterile_norm}	    
  \end{subfigure}
\caption{Shows the effect of varying either $M_{A}^{eff}$ or $\Delta m^{2}$ on the predicted reconstructed energy distribution. Both shape only (area normalised to unity), and normalised plots are shown. The MiniBooNE data points are shown on the normalised plots for comparison, the size of the errors on these points indicates the strong correlations between reconstructed energy bins.}\label{fig:varyParams}
\end{figure}

The plots in Figure~\ref{fig:varyParams} provide a visual confirmation that the analysis method and cross section model used in this analysis produce sensible values for the event rate in MiniBooNE. Figures~\ref{subfig:NCEL_ma_shape} and~\ref{subfig:NCEL_ma_norm} illustrate the effect that varying the single free cross section parameter, $M_{A}^{eff}$, has on the predicted NCEL distribution. It can be seen from Figure~\ref{subfig:NCEL_ma_shape} that an increasing value of $M_{A}^{eff}$ only has a large effect on the shape of the distribution at low values of $T_{reco}$, though it can be seen in Figure~\ref{subfig:NCEL_ma_norm} that the overall normalisation increases with increasing $M_{A}^{eff}$.

Because there are more variable sterile parameters, it is difficult to illustrate the effect that sterile parameters can have on the distribution. Figures~\ref{subfig:NCEL_sterile_shape} and~\ref{subfig:NCEL_sterile_norm} show the effect that different values of $\Delta m^{2}_{42}$ have on the predicted distributions; the other parameters have been fixed for simplicity. $U_{\mu 4} = 0.4$ has been chosen because it is around the limit placed by an analysis of atmospheric neutrinos for all values of $\Delta m^{2}$~\cite{giunti11Conf, maltoni07}, the other independent parameters, $U_{e4} = U_{\tau 4} = 0.2$ have been chosen to be equal for simplicity and small to keep the $U_{s4}$ component large, as would be expected. These example sterile parameters correspond to $\sin^{2}2\vartheta_{\mu s} \approx 0.49$.

As all of the sterile parameters affect the $T_{reco}$ distribution, the relationship between any single sterile model parameter and the $T_{reco}$ distribution is hard to visualise. Increasing $U_{\mu 4}$ will reduce the NCEL signal, but for any value of $U_{\mu 4}$, increasing values of $U_{e4}$ and $U_{\tau 4}$ will increase the NCEL signal. The sterile oscillations decrease the event rate more in low $T_{reco}$ bins, as seen in Figure~\ref{subfig:NCEL_sterile_norm}, causing a subtle shift in the shape across the entire distribution as can be seen in Figure~\ref{subfig:NCEL_sterile_shape}. However, the effect on shape is complicated by the $\nu_{e}$ contamination in the beam, which causes differences in the way $U_{e4}$ and $U_{\tau 4}$ affect the shape (because the shape of the $\nu_{e}$ flux is not the same as the $\nu_{\mu}$ flux~\cite{mbFlux}). But the effect of the $\nu_{e}$ contamination on the shape is minimal as the contamination is only 0.52\% of the total flux~\cite{mbFlux}.

\section{NCEL fits}
\label{sec:ncelFit}
\subsection{Fitting procedure}
The best fit points are obtained by minimising the chi-square statistics defined in Equation~\ref{eq:ncelChiSq_sim} and Equation~\ref{eq:ncelChiSq_seq}, where $\mathbf{\theta}$ are the parameters which are minimised in the fit, $M^{-1}_{ij}$ is the covariance matrix published with~\cite{mbNCEL, PerevalovThesis} and $i, j$ are reconstructed energy bins. The minimisations were performed using the MINIMIZE algorithm (MIGRAD algorithm, reverting to the SIMPLEX algorithm if there is no convergence) in the MINUIT minimiser~\cite{minuit} within the ROOT framework~\cite{root}. The IMPROVE algorithm was used several times (alternating with calls to MINIMIZE) to ensure that the minimum in each case was global rather than local.
\begin{equation}
\indent
\chi^{2} (\mathbf{\theta}) = \sum^{51}_{i=0}\sum^{51}_{j=0} \left(\nu_{i}^{DATA} - \nu_{i}^{MC}(\mathbf{\theta}) \right) M_{ij}^{-1} \left(\nu_{j}^{DATA} - \nu_{j}^{MC}(\mathbf{\theta}) \right)
\label{eq:ncelChiSq_sim}
\end{equation}
Equation~\ref{eq:ncelChiSq_sim} is used in the simultaneous fit, where the free parameters $\mathbf{\theta}$ are $\Delta m^{2}$, $U_{e4}$, $U_{\mu 4}$, $\sin^{2}2\vartheta_{\mu s}$ and $M_{A}^{eff}$. It is also used in the $M_{A}^{eff}$ only fit, where all of the sterile parameters are set to zero.
\begin{equation}
\indent
\chi^{2} (\mathbf{\theta}) = \sum^{51}_{i=0}\sum^{51}_{j=0} \left(\nu_{i}^{DATA} - \nu_{i}^{MC}(\mathbf{\theta}) \right) M_{ij}^{-1} \left(\nu_{j}^{DATA} - \nu_{j}^{MC}(\mathbf{\theta}) \right) + \left(\frac{\mathbf{\theta}_{M_{A}}}{\sigma_{M_{A}}} \right)^{2}
\label{eq:ncelChiSq_seq}
\end{equation}
Equation~\ref{eq:ncelChiSq_seq} is used in the sequential fit, where the additional penalty term uses the one sigma error on $M_{A}^{eff}$, $\sigma_{M_{A}}$, obtained in the $M_{A}^{eff}$ only fit. It should be kept in mind that the sequential fit is only statistically rigorous if the cross section and sterile neutrino parameters are completely uncorrelated, if there are correlations, this procedure will give incorrect results.

As shown in Section~\ref{sec:sterileModel}, the value of $\sin^{2}2\vartheta_{\mu s}$ depends on $U_{\mu4}$ and $U_{s4}$, or equivalently on $U_{e4}$, $U_{\mu4}$ and $U_{\tau 4}$ (given the unitarity constraint). The parameter $U_{s4}$ cannot be measured directly as the NCEL measurement is not made in a pure $\nu_{\mu}$ beam, so the latter combination must be used. This leaves a 4 dimensional sterile parameter space to scan, which would be very expensive computationally. Instead, the $\Delta m^{2} - \sin^{2}2\vartheta_{\mu s}$ plane is scanned, and the other sterile parameters are allowed to vary to minimise the chi-square but whilst also obeying the unitarity constraint and the constraint imposed by fixing $\sin^{2}2\vartheta_{\mu s}$. The unitarity constraint is enforced by including a penalty term in the chi-square, forcing the fitter into the physically allowed region. Although MIGRAD relies on calculating derivatives, and as such could have problems with these discontinuities, the use of the SIMPLEX 
algorithm (which does not calculate derivatives) if MINUIT failed helped to guide the fitter away from problem regions.

\subsection{$M_{A}^{eff}$ fit}
\label{sec:ncelMAFit}
The fit to $M_{A}^{eff}$ serves two purposes. As all of the sterile parameters are set to zero, it gives the chi-square value of the null hypothesis. It is also used as the cross section measurement in the sequential fit, providing a penalty term on the value of $M_{A}^{eff}$. The error on $M_{A}^{eff}$ is calculated by moving the $M_{A}^{eff}$ value away from the best fit incrementally until $\Delta \chi^{2} = 1$.
\begin{table}[h]
  \centering
    \begin{tabulary}{1.0\textwidth}{CCCC}
	  \hline\hline
	  & {\bf $\chi^{2}$} & {\bf M$_{A}$} & {\bf DOF} \\\hline\hline
	  This analysis & 32.060 & 1.240 $\pm$ 0.076 & 50 \\
	  MiniBooNE~\cite{mbNCEL} & 26.9 & 1.39 $\pm$ 0.11 & 50 \\
	  \hline\hline	
      \end{tabulary}
  \caption{Best fit values for the M$_{A}$ only fit to the NCEL sample, along with the published MiniBooNE value for comparison.}	
  \label{tab:maOnlyNCEL}
\end{table}

Table~\ref{tab:maOnlyNCEL} shows the best fit value of $M_{A}^{eff}$ found in this analysis, along with the calculated error. For comparison, the published MiniBooNE result~\cite{mbNCEL} is included. The value of $M_{A}^{eff}$ found in this analysis is lower than the published MiniBooNE result, probably due to differences in the generators used. The enhanced Pauli blocking in the MiniBooNE cross section model and the different values of $\Delta s$ between the generators have both been shown to have an effect on the calculated $M_{A}^{eff}$ value~\cite{mbNCEL, PerevalovThesis}. Indeed, in the MiniBooNE paper, there is a measurement of $\Delta s$ using the ratio of $\nu p \rightarrow \nu p$ to $\nu N \rightarrow \nu N$ as a function of reconstructed nucleon energy~\cite{mbNCEL}. They find a value of $\Delta s = 0.00 \pm 0.30$ for $M_{A} = 1.23$ GeV, which they note is consistent with a previous measurement by BNL E734~\cite{e734NCEL}, and is consistent with the value of $M_{A}^{eff}$ found here given that the 
E734 value of $\Delta s$ 
was used.

\subsection{Best fit results}
Table~\ref{tab:bestFitsNCEL} gives the best fit values for both the sequential and simultaneous fits. It is interesting that the best fit values are very different between sequential and simultaneous fits, indicating that there are correlations between the cross section and sterile model parameters. This highlights how the sequential fit method could mask a sterile signal - a low value of $M_{A}^{eff}$ could compensate for disappearance in the signal due to sterile oscillations, masking the disappearance in the subsequent fit to the sterile parameters. It is also interesting that $M_{A}^{eff}$ tends to a much higher value in the simultaneous fit, much higher than is expected.

\begin{table}[h]
  \centering
    \begin{tabulary}{1.0\textwidth}{CCC}
	  \hline\hline
	                           & Sequential & Simultaneous \\\hline\hline
	  $\chi^{2}$               & 27.717 & 23.684 \\
	  $\Delta m^{2}$           & 5.904  & 2.588  \\
	  U$_{e4}$                 & 0.570  & 0.474  \\
	  U$_{\mu4}$               & 0.707  & 0.745  \\
	  sin$^{2}2\vartheta_{\mu s}$ & 0.349  & 0.490  \\
	  M$_{A}$                  & 1.307  & 1.714  \\
	  DOF                      &  47    &  46    \\
          \hline\hline	
      \end{tabulary}
  \caption{Best fit values for the NCEL fits.}	
  \label{tab:bestFitsNCEL}
\end{table}

The lowest values found during the parameter scans were used as initial values when calculating the best fit points. This reduced the computation time, and ensured that the fits did not become trapped in local minima as sometimes happened when fits were performed using randomly generated starting values for all parameters.

\subsection{Parameter scans}
\label{sec:NCELpScans}
Chi-square values for 9000 points in the $\Delta m^{2}-\sin^{2}\vartheta_{\mu s}$ plane were calculated, with 120 $\Delta m^{2}$ points distributed logarithmically in the region $0.1 \leq \Delta m^{2} \leq 10$ eV$^{2}$ and 75 $\sin^{2}\vartheta_{\mu s}$ points in the region $0.005 \leq \sin^{2}\vartheta_{\mu s} \leq 0.745$ with spacing $\delta\sin^{2}\vartheta_{\mu s} = 0.01$. The confidence regions are calculated using the constant $\Delta \chi^{2}$ method, $\chi^{2}_{allowed} \leq \chi^{2}_{min} + \Delta \chi^{2}$, where the best fit value $\chi^{2}_{min}$ is given in Table~\ref{tab:bestFitsNCEL}, and $\Delta \chi^{2}$ is calculated for 2 degrees of freedom: 4.61 - 90\% confidence level; 9.21 - 99\% confidence level~\cite{pdg2012}.

\begin{figure}[h!]
 \centering
 \begin{subfigure}[htb]{0.49\textwidth}
    \includegraphics[width=\textwidth]{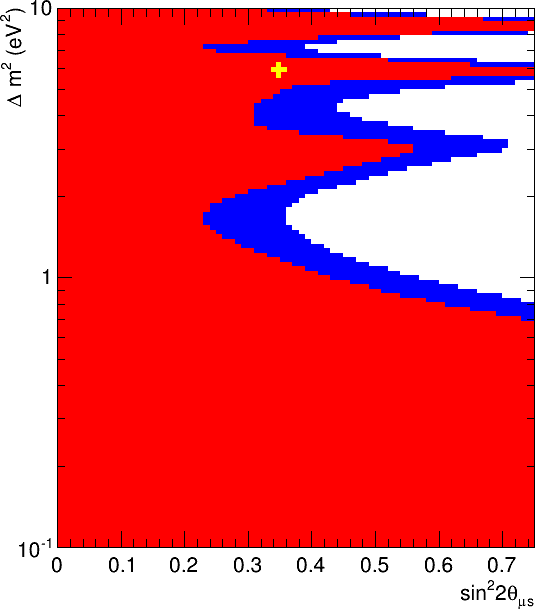}
    \caption{NCEL sequential fit}
    \label{subfig:pScan_NCEL_seq}	
 \end{subfigure}
 \begin{subfigure}[htb]{0.49\textwidth}
    \includegraphics[width=\textwidth]{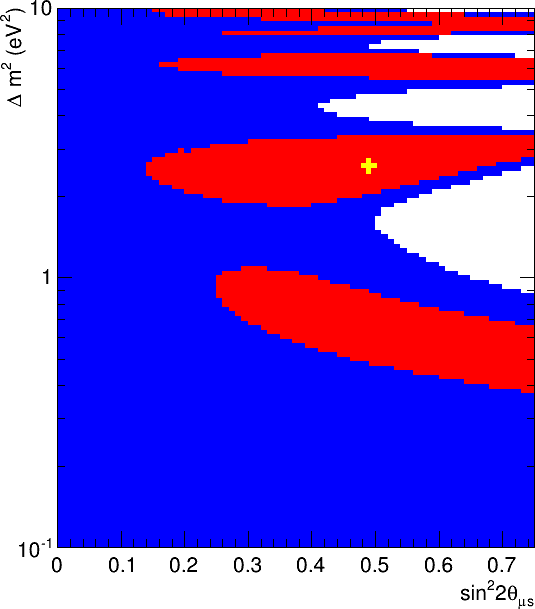}
    \caption{NCEL simultaneous fit}
    \label{subfig:pScan_NCEL_sim}	    
 \end{subfigure}
\caption{The exclusion plots produced by both the sequential and simultaneous fit techniques for the MiniBooNE NCEL dataset. The 90\% region is shown in red, the 99\% region is shown in blue, and the best fit point is indicated with a yellow cross.}\label{fig:pScansNCEL}
\end{figure}

\begin{figure}[h]
 \centering
 \begin{subfigure}[htb]{0.49\textwidth}
    \includegraphics[width=\textwidth]{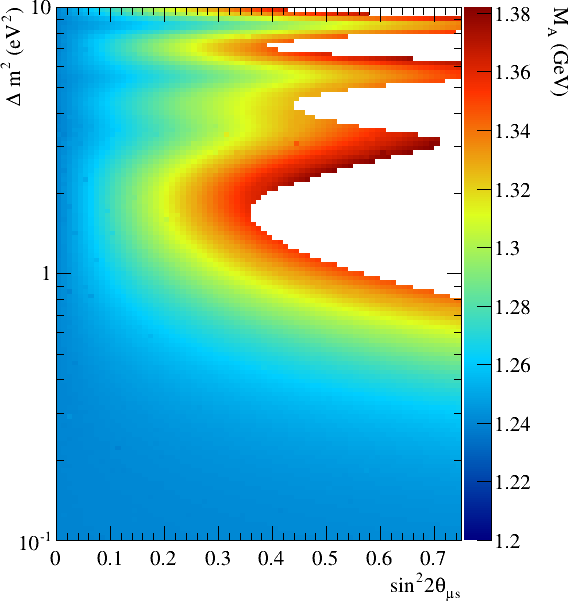}
    \caption{NCEL sequential fit}
    \label{subfig:pScan_NCEL_seq_Ma}	
 \end{subfigure}
 \begin{subfigure}[htb]{0.49\textwidth}
    \includegraphics[width=\textwidth]{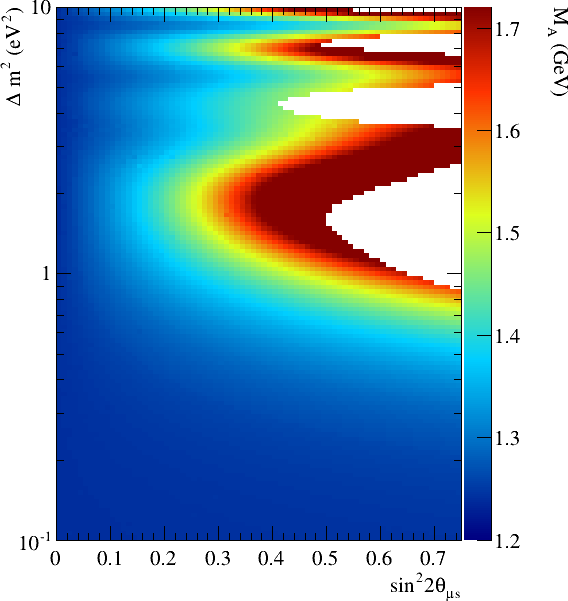}
    \caption{NCEL simultaneous fit}
    \label{subfig:pScan_NCEL_sim_Ma}	    
 \end{subfigure}
\caption{The variation in the best fit value for $M_{A}^{eff}$ across the 99\% region in the $\Delta m^{2}-\sin^{2}\vartheta_{\mu s}$ plane.}\label{fig:pScansNCEL_Ma}
\end{figure}

The allowed regions for the sequential fit are shown in Figure~\ref{subfig:pScan_NCEL_seq}, and for the simultaneous fit in Figure~\ref{subfig:pScan_NCEL_sim}. The variation in the best fit values for $M_{A}^{eff}$ across the 99\% allowed regions is shown in Figure~\ref{subfig:pScan_NCEL_seq_Ma} for the sequential fit, and in Figure~\ref{subfig:pScan_NCEL_sim_Ma} for the simultaneous fit. Although the best fit value of the simultaneous fit is high, this is not the case for much of the allowed region.

\section{Discussion and conclusions}
\label{sec:conclusion}
The $M_{A}^{eff}$ only fit, shown in Section~\ref{sec:ncelMAFit}, gave a value, $M_{A}^{eff} = 1.240 \pm 0.076$ GeV, which is consistent with the published MiniBooNE result of $M_{A}^{eff} = 1.39 \pm 0.11$ GeV; the differences between the two values can be understood in terms of the slight differences between the cross section models, and the different generators used. This is a useful sanity check for the method used to produce event rate predictions for this analysis.

Two fits to a 3+1 sterile neutrino model were performed, the sequential fit, which mimics previous MiniBooNE $\nu_{\mu}$-disappearance analyses~\cite{mbDisApp09, mbSciNu2012, mbSciaNu2012} by implicitly assuming that the cross section and sterile neutrino model parameters are uncorrelated. And the simultaneous fit, where all parameters are fit concurrently, making no assumption about the correlations between models. Given the current uncertainty surrounding neutrino cross section predictions, discussed in Section~\ref{sec:intro}, it is not possible to use constraints on $M_{A}^{eff}$ from other experiments as the effective axial mass is so dependent on experimental details. Until this uncertainty is resolved, the only consistent way to produce short baseline sterile neutrino limits is to perform a sequential or simultaneous fit as described here (note that this is not the case if there is a near detector where oscillations can be neglected). We find that the sequential and simultaneous fits produce different 
best fit values and contours, as can be seen in Figure~\ref{fig:varyParams}. This shows that for the NCEL dataset, it is wrong to assume that the sterile and cross section model parameters are uncorrelated. As such, it should be stressed that the sequential fit shown here is not correct.

It is, however, interesting to compare the contours produced by sequential and simultaneous fits. The sequential fit produced stronger limits in the $\Delta m^{2}-\sin^{2}\vartheta_{\mu s}$ plane as would be expected if the sterile and cross section model parameters are correlated but not treated as such in the fit. The cross section parameters are pulled so as to partially mask a signal, or a statistical fluctuation that mimics a signal. Limits produced by sequential fits should be therefore be treated with caution unless it is shown that there are no correlations between models.

The 90\% and 99\% confidence regions produced by the simultaneous fit are shown in Figure~\ref{subfig:pScan_NCEL_sim}. These are the main result of this analysis and are the first short baseline oscillation result in the $\Delta m^{2}-\sin^{2}\vartheta_{\mu s}$ plane. The 99\% limits produced by this analysis are not particularly strong as a result of the freedom between the sterile mixing parameters $U_{e4}$, $U_{\mu 4}$ and $U_{\tau 4}$ - a large change in one value can be countered by large changes in the others to diminish the effect on the signal. Much stronger limits can be produced by performing a joint fit to the NCEL and MiniBooNE CCQE cross section measurement~\cite{mbCCQE, KatoriThesis}, which provides an additional constraint on $U_{\mu 4}$. However, as the covariance matrix was not included in the public data release for the CCQE measurement, and as there is insufficient information available to properly account for correlated systematics between the samples, this fit has not been included in 
this paper, though it can be found in~\cite{myThesis}. This analysis does find that the 3+1 model is favoured over no oscillations to greater than 90\% confidence, which is an intriguing result, however the best fit point tends towards a value of $M_{A}^{eff}$ which is considerably higher than is found by other experiments~\cite{zeller12}, though it can be seen in Figure~\ref{fig:pScansNCEL_Ma} that $M_{A}^{eff}$ is not as high for much of the allowed regions. The mass splitting of $\Delta m^{2} = 2.588$ eV$^{2}$ at the best fit point, is in conflict with global best fit values for 3+1 mixing models~\cite{kopp13GF, conrad12GF, abazajian12}.

Although the 90\% contours are interesting, we feel that it is prudent to sound a note of caution. There are two possible issues for this and other sterile analyses which may cause these differences. First, it is possible that the NCEL dataset is insufficient to constrain both the cross section and sterile neutrino model parameters, however work fitting both datasets suggests that this is not the cause~\cite{myThesis}. Second, it is possible that the differences between this analysis and $\nu_{\mu}$-disappearance analyses are caused by the inadequacies of the RFG model. Here we followed the assertion made in~\cite{mbAntiNCEL} that an inflated $M_{A}^{eff}$ is a reasonable, though ad hoc, way to model the additional multi-nucleon effects. If this is not the case, the differences between multi-nucleon contributions will affect the sterile neutrino fit, and this effect may not be the same for NCEL and CCQE selections, which could explain the different preferred values for the sterile parameters found through 
sterile fits to these datasets. 
\begin{figure}[h!]
 \centering
  \includegraphics[width=0.6\textwidth]{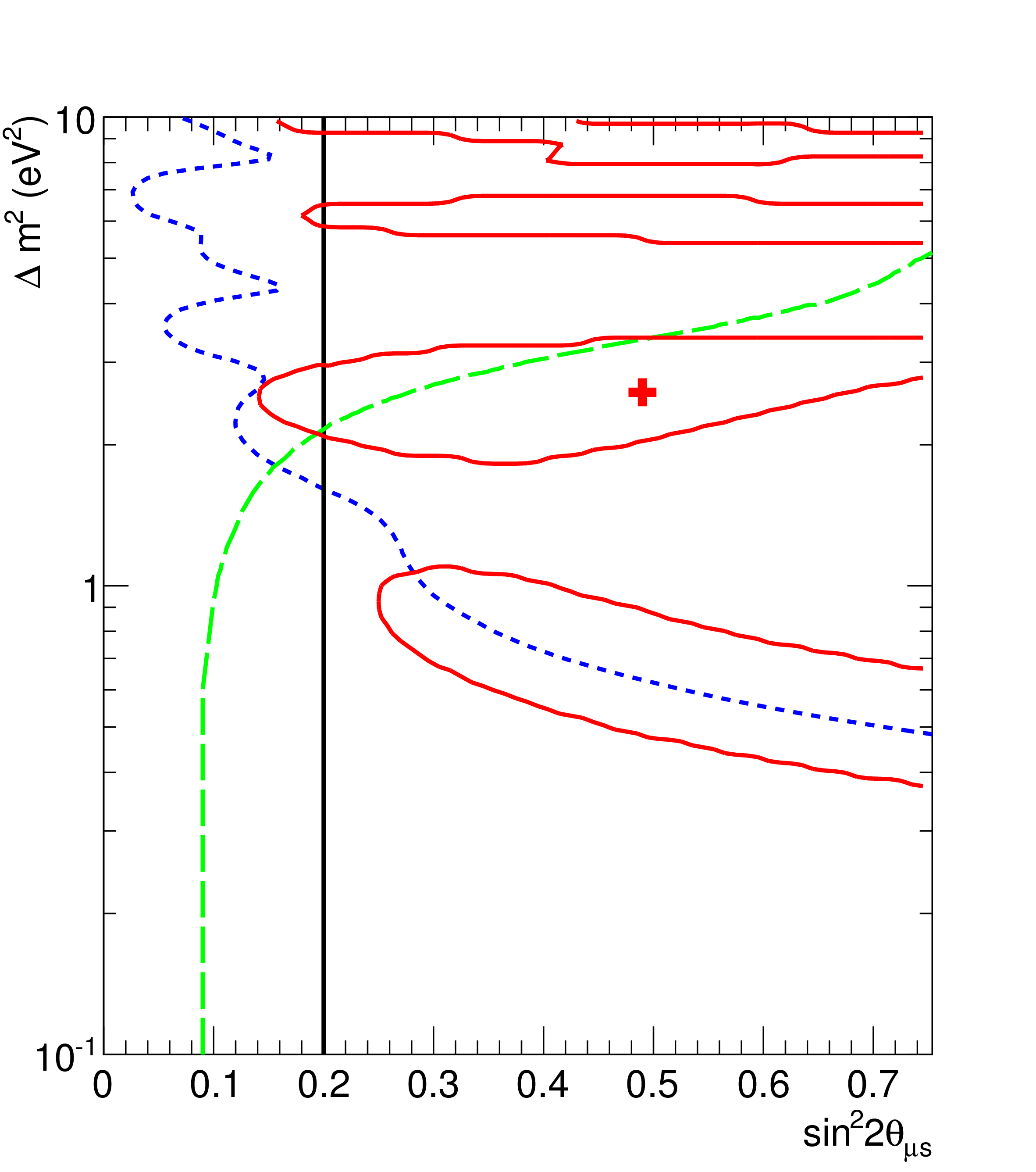}
  \caption{The 90\% confidence region from the simultaneous fit is shown (solid red line), with the best fit point indicated by the red cross. Also shown are limits from other experiments: MiniBooNE-SciBooNE $\nu_{\mu}$-disappearance limits using the spectral fit method~\cite{mbSciNu2012} (short dashed blue line); limits from the analysis of atmospheric data~\cite{maltoni07} (black solid line); limits extracted in~\cite{giunti11MINOS} from the MINOS NC-disappearance analysis~\cite{minos11} (long dashed green line). The authors of~\cite{giunti11MINOS} consider oscillations in the MINOS near detector to set limits over a wider range of $\Delta m^{2}$ values using a two-parameters least-squares analysis, the limit given here is approximate as it is taken from the plot in the paper (Figure 6).}\label{fig:sterileComparison}	    
\end{figure}

The comparison with other published sterile neutrino limits shown in Figure~\ref{fig:sterileComparison} highlights the disagreement with other datasets. Note that limits on $\sin^{2}2\vartheta_{\mu\mu}$ have been treated as if they are limits on $\sin^{2}2\vartheta_{\mu s}$ in Figure~\ref{fig:sterileComparison}, this is justified because $\sin^{2}2\vartheta_{\mu s} \leq \sin^{2}2\vartheta_{\mu\mu}$. The MINOS NC limit~\cite{giunti11MINOS, minos11, minos10} is only a strong constraint for a small range of $\Delta m^{2}$ because possible oscillations at the near detector weaken the limit, but their 90\% limit excludes the best fit point we find in this analysis and some of our 90\% allowed region. The MiniBooNE-SciBooNE limit depends implicitly on the value of $M_{A}^{eff}$ measured by the experiment, however MiniBooNE assert in~\cite{mbDisApp09} that the value of $M_{A}^{eff}$ is uncorrelated with the sterile model parameters. We feel that this is the most interesting comparison, as the difference between the 
NCEL and CCQE sterile analyses may point to a problem with the cross section model. The atmospheric constraint 
alone rules out much of the 90\% preferred region in this analysis, a recent reanalysis of this constraint by the Super-Kamiokande collaboration presented in a recent conference talk~\cite{himmel13} is even stronger than that found in~\cite{maltoni07}, finding $\sin^{2}2\vartheta_{\mu\mu} \leq 0.131$ to 99\% confidence, which conflicts with all of the 90\% parameter space found in this analysis.

There are strong bounds on U$_{e4}$ from reactor experiments (for a summary of reactor constraints, see ~\cite{conrad12GF, kopp13GF}), which are not accounted for in this analysis. However, changes to U$_{e4}$ can be almost fully compensated for by changes in U$_{\tau 4}$. The only difference arises from the effect U$_{e4}$ has on the small amount of $\nu_{e}$ contamination in the beam (less than 0.52\% of the total flux~\cite{mbFlux}). Therefore including reactor constraints to the fit performed here would only have a minimal effect on the chi-square value found at each fitted point, though the value for U$_{\tau 4}$ would increase and the value for U$_{e4}$ would decrease. Strong constraints on both U$_{\tau 4}$ and U$_{e4}$ would, however, affect the contours found in this analysis, which should be kept in mind if these results are used in a global fit.

A future deeper understanding of the underlying neutrino cross sections, which models the current inconsistencies well, would provide reliable and independent cross section measurements which sterile neutrino experiments can use when placing limits. However, the sterile neutrino limits are dependent on the cross section model used to make event rate predictions, and as such any limits produced under the assumption of RFG models need to be treated in the sterile neutrino literature with the same wariness that RFG models are in the cross section literature. When a consistent model describing neutrino cross sections at these energies has emerged, it will be interesting to see whether reanalysis of existing sterile datasets produces significantly different limits.

\section*{Acknowledgements}
We would like to thank Jonathan Perkin, Matthew Lawe and Leon Pickard for useful discussions and comments on all aspects of this work. C.W. would like to thank Denis Perevalov for his help in using the information contained in Appendix B of~\cite{PerevalovThesis} to produce a response matrix, and for useful discussions in the early stages of this work. C.W. would also like to thank Thomas Dealtry for useful discussions on how to best use GENIE~\cite{genieMC}. We would like to thank the STFC for the ongoing funding of the neutrino group at the University of Sheffield, C.W. would also like to thank the STFC for his PhD Studentship which supported this work.

\bibliographystyle{JHEP}
\bibliography{steriles}

\end{document}